\documentclass{iau_FM}

\def \msun{\ifmmode{{\rm\ M}_\odot}\else{${\rm\ M}_\odot$}\fi}

\usepackage{graphicx}

\title[Type II supernova diversity] 
{Type II supernova diversity}

\author[Joseph Anderson]   
{Joseph P. Anderson$^1$}

\affiliation{$^1$European Southern Observatory,
Alonso de Córdova 3107, Casilla 19001,
Santiago, Chile \\email: {janderso@eso.org}}

\pubyear{2015}
\jname{Astronomy in Focus, Volume 2} 
\editors{Piero Benvenuti, ed.}
\begin{document}

\maketitle

\begin{abstract}
It is now firmly established that at a significant fraction of 
hydrogen-rich type II supernovae (SNe II) arise from red supergiant 
progenitors. However, a large diversity of SN properties exist, 
and it is presently unclear how this can be understood 
in terms of progenitor differences and pre-SN stellar evolution. In 
this contribution, I present the diversity of SN II $V$-band light-curves 
for a large sample of SNe~II, and compare these to photometry of SNe~II
which have progenitor mass constraints from pre-explosion imaging. 
\keywords{(stars:) supernovae: general}
\end{abstract}

Type II supernovae (SNe~II) are the explosions of massive ($>$8-10\msun) stars
which have retained a significant fraction of their hydrogen envelopes.
Historically, these events were separated into `Plateau' IIPs which
show almost constant luminosity for $\sim$100 days, and `Linear' IIL 
which decline much faster in a linear manner (\cite[Barbon, Ciatti \&\ Rosino 1979]{bar79}).
However, recent work has questioned this separation and argued for 
a continuum of events with very few (if any) SNe showing the historically defined
morphology of a SN~IIL (\cite[Anderson et al. 2014]{and14}; \cite[Sanders et al. 2014]{san14}). 
The direct identification of
red supergiant (RSG) progenitors on pre-explosion images (e.g. \cite[Smartt 2015]{sma15}) 
is concrete evidence that SNe~II arise
from the explosions of RSG stars. However, it is unclear how
the large diversity of SN~II light-curves and spectra is linked to 
progenitor properties such as mass, metallicity, extent of the pre-explosion
hydrogen envelope, or the role of binary interactions.
Analysing the diversity within large samples of events, and searching for correlations between 
parameters to try to understand the whole ensemble of explosions through
consistent analyses can aid in our understanding. This was recently achieved in Anderson et al. (2014, A14), and
this contribution I briefly summarise some of the important results from that work, while
comparing those light-curves to the same photometric measurements for SNe~II with progenitor mass constraints.\\

\begin{figure}
\centering
\includegraphics[width=6.5cm]{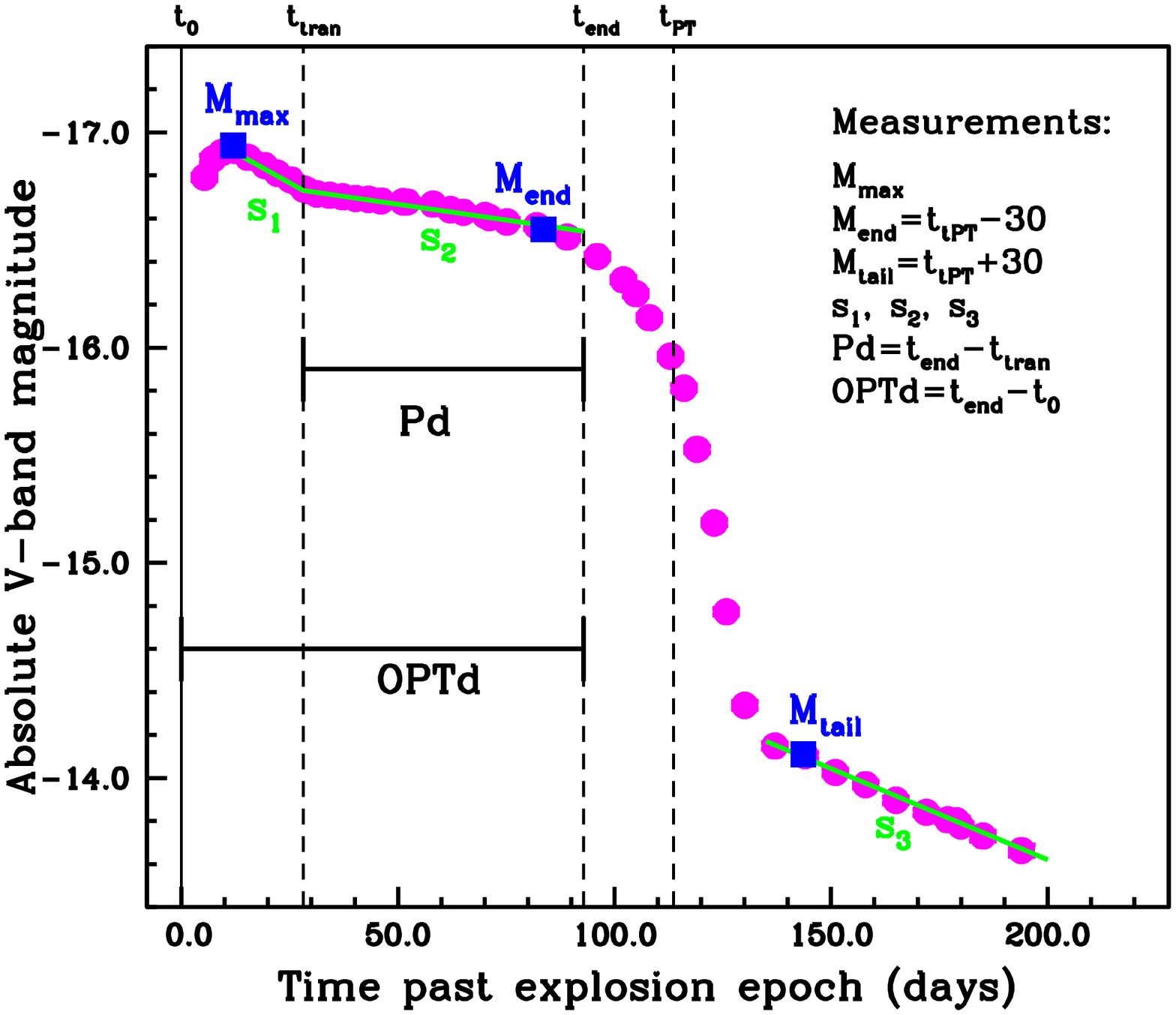} 
\includegraphics[width=6.5cm]{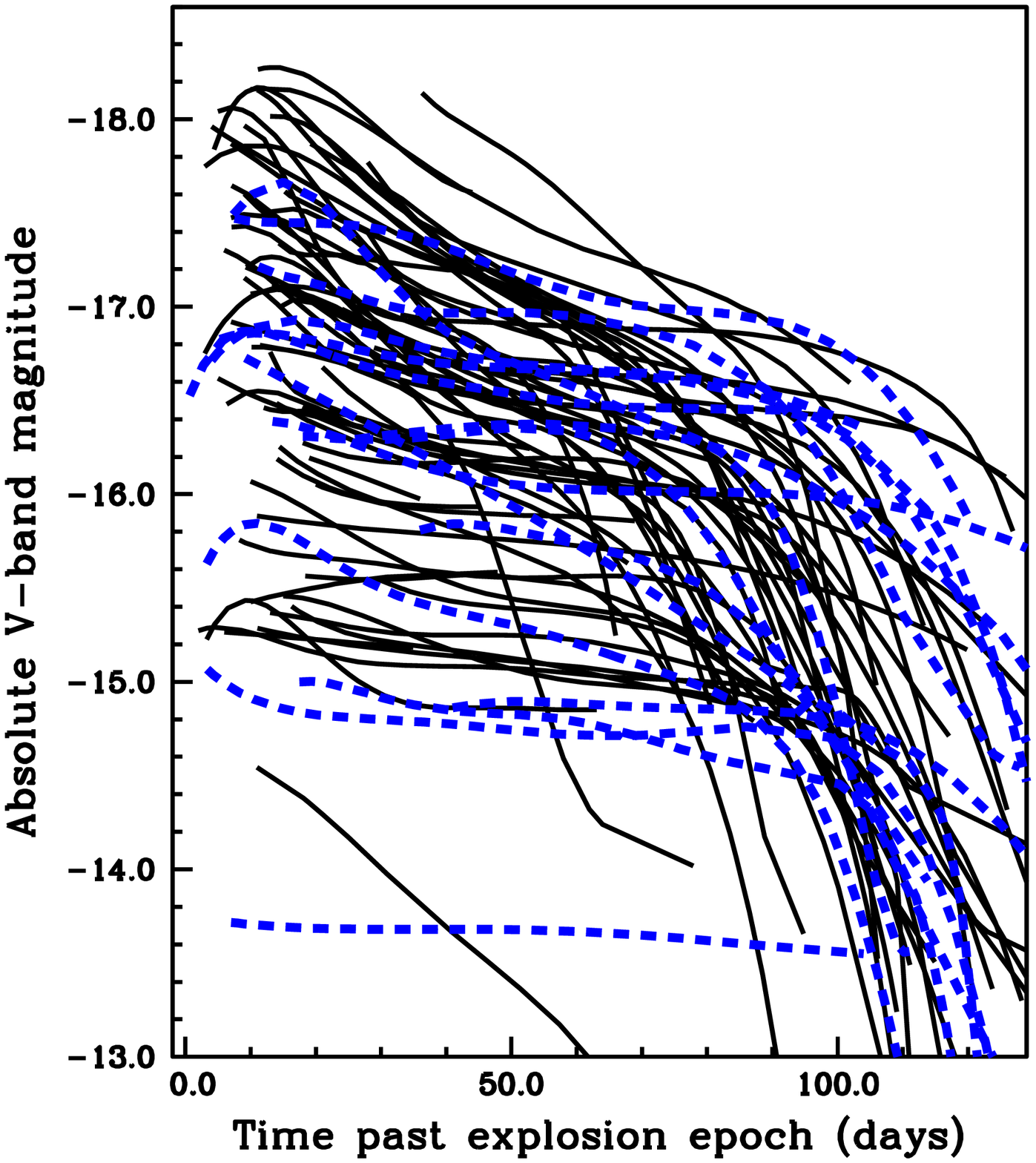} 
\caption{\textbf{\textit{Left, a)}}: SN~II light-curve schematic outlining
the $V$-band photometric parameters defined and measured for all SNe~II in this
work. \textbf{\textit{Right, b)}}: absolute $V$-band light-curves of SNe~II. Those in solid lines
are from A14, while those in dashed lines are SNe~II with direct progenitor
detection constraints in the literature.}
\end{figure}

\indent In Fig.~1a I present a schematic outlining the $V$-band light-curve parameters
chosen in A14 to characterise the photometric diversity of
SNe~II. These include a range of absolute magnitudes at different phases; a range of decline rates; 
and various time epochs and time durations. Some of these parameters were previously 
well documented, however others were not (see A14 for a detailed discussion). 
In particular, the early decline rates ($s_1$ in our nomenclature)
have received little discussion in the literature. In A14 we also highlight differences
between the optically thick phase duration (OPTd, historically named the plateau duration): 
the time between explosion and the end of the plateau, and the plateau duration (Pd) which
we now define as the time duration for which a SN is actually showing a `plateau', i.e. from the
end of the initial $s_1$ decline to the end of  $s_2$ when the SN 
starts transitioning to the later radioactively powered phase.\\
\indent All photometric parameters defined in Fig.~1a were measured (where possible) for the
full sample of $>$100 SNe~II in A14. The same parameters are also measured for all those SNe~II with progenitor
mass constraints. In Fig.~1b the absolute $V$-band magnitude light-curves
for both the A14 and direct detection samples are presented. This figure clearly shows the large 
diversity seen in SN~II light-curves. Diversity is observed in terms of brightness, decline
rates, and optically thick phase durations. The direct detection sample
also spans a wide range of light-curve morphologies. This is interesting, especially in the
context of their distribution of progenitor mass constraints. Progenitors 
appear to have a relatively small range in progenitor mass, 
between 8 and 17\msun\ (Smartt 2015). It is thus pertinent to ask: can such
observed light-curve diversity be explained by progenitors of such similar
mass?

\begin{figure}
\centering
\includegraphics[width=6.5cm]{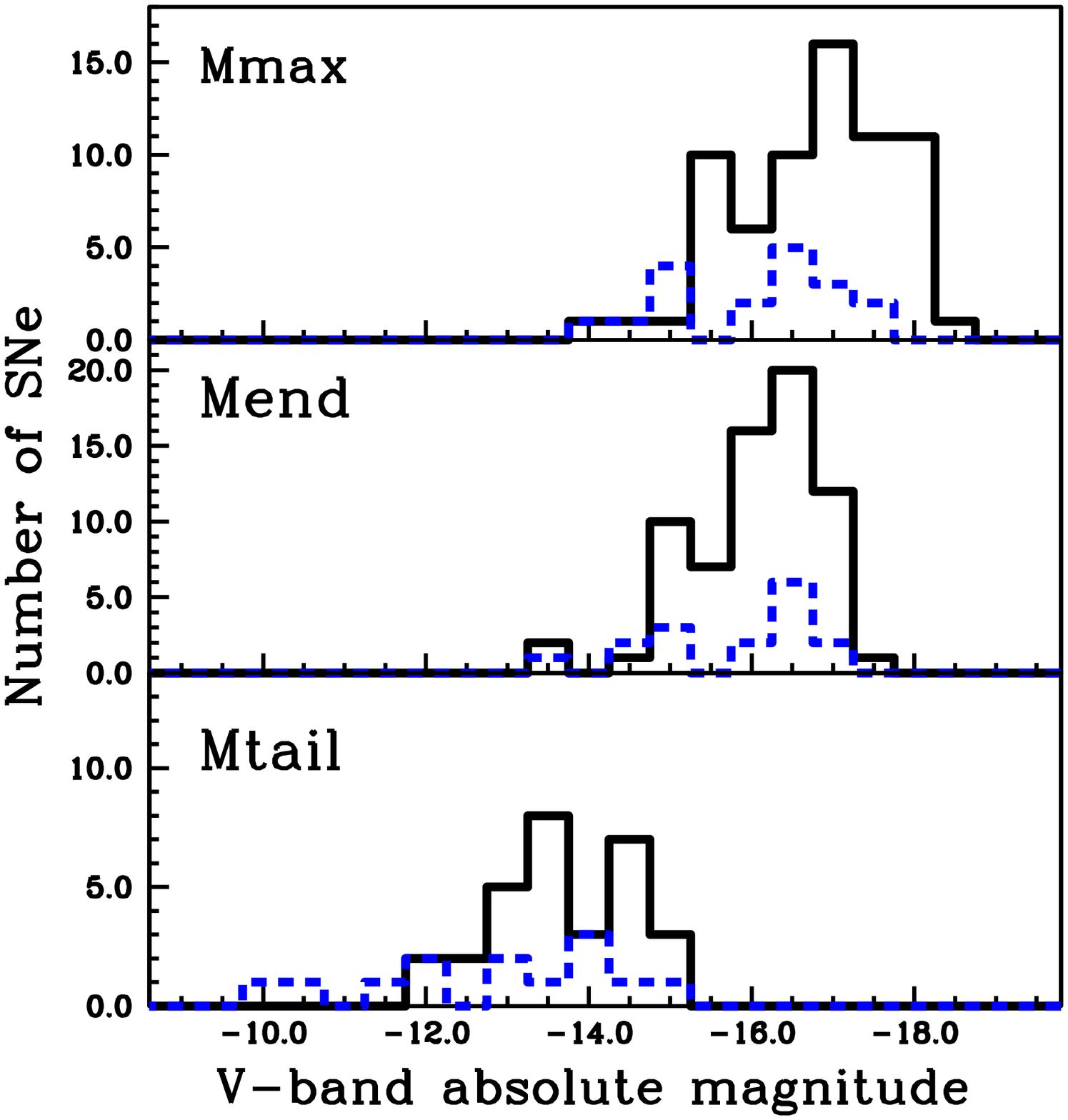} 
\includegraphics[width=6.5cm]{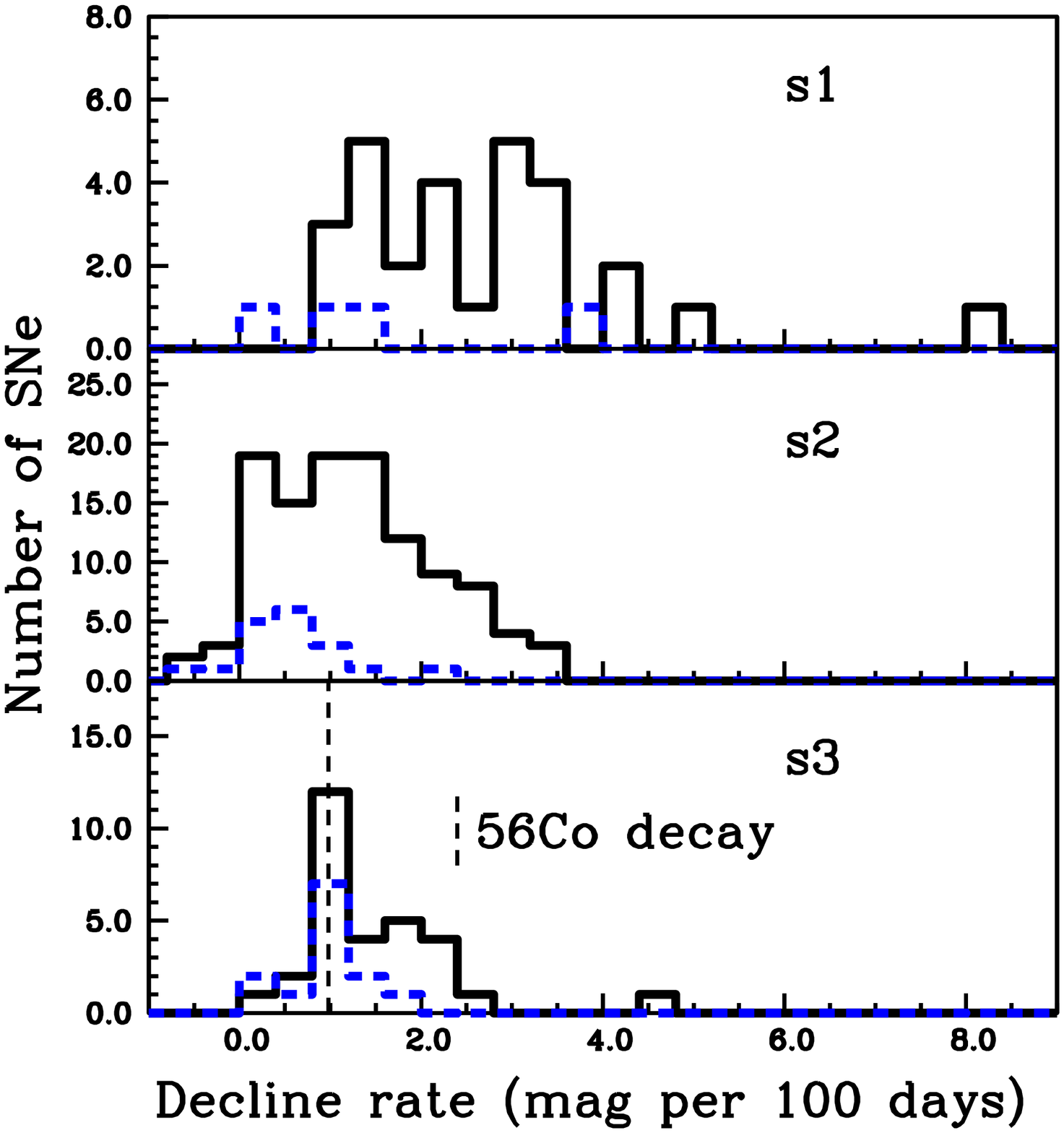} 
\caption{\textbf{\textit{Left, a)}}: Histograms of the three absolute 
magnitudes of SNe~II. \textbf{\textit{Right, b)}}: Histograms of the three decline
rates of SNe~II. Those in solid lines
are from A14, while the dashed lines are SNe with progenitor
detection constraints.}
\end{figure}

\indent In Fig.~2 histograms are presented of SN~II absolute magnitudes and decline rates.
The A14 sample is on average brighter than the direct detection sample. This is easily explained
through the fact that the A14 sample is magnitude limited, while the direct detection is
essentially a volume limited sample. With respect to decline rates, it is seen that there are
very few direct detections that have measured $s_1$ declines.
Fig.~2b shows that there is a wide range of initial decline rates, and understanding
this early behaviour could have strong constraints on SN~II physics.
The $s_2$ distribution of the direct detections is shifted towards lower values, and this
is a direct consequence of the correlation between $M_{max}$ and $s_2$ as presented in Fig.~3.
Another interesting feature of Fig~2b is the distribution of $s_3$ values. $s_3$ is the decline rate during the
radioactively powered phase. Assuming full trapping of gamma-rays
one expects SNe to follow a standard decline rate at these epochs. However, significant deviation
from this expectation is observed. This hints at some SNe~II having ejecta masses insufficient in mass
and/or density to fully trap the emission.\\
\indent In Fig.~3a I present $M_{max}$ against $s_2$, while in Fig.~3b $M_{max}$ against
OPTd. A correlation is observed in that brighter SNe decline more quickly. A large
range in OPTd values is also found, suggesting a large range in hydrogen envelope masses 
at the explosion epoch. Finally, I note that in all distributions --both those presented here and
those in A14 and \cite[Guti\'errez et al. (2014)]{gut14}-- an observational continuum
is seen.
Further statistical studies of this kind will deepen our knowledge of the SN~II phenomenon and 
strengthen our confidence in using these explosions to probe other processed in the Universe.

\begin{figure}
\centering
\includegraphics[width=6.5cm]{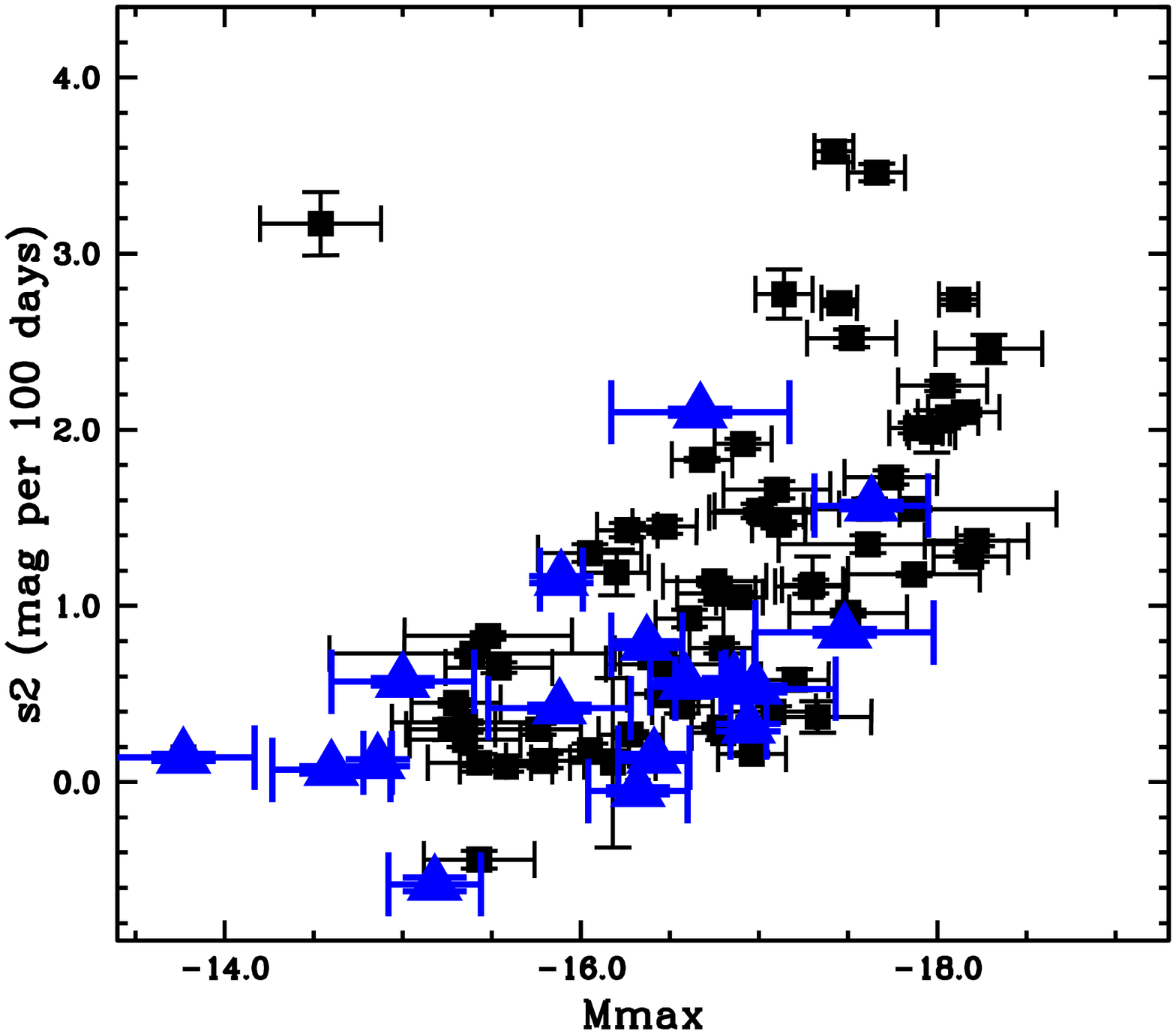} 
\includegraphics[width=6.5cm]{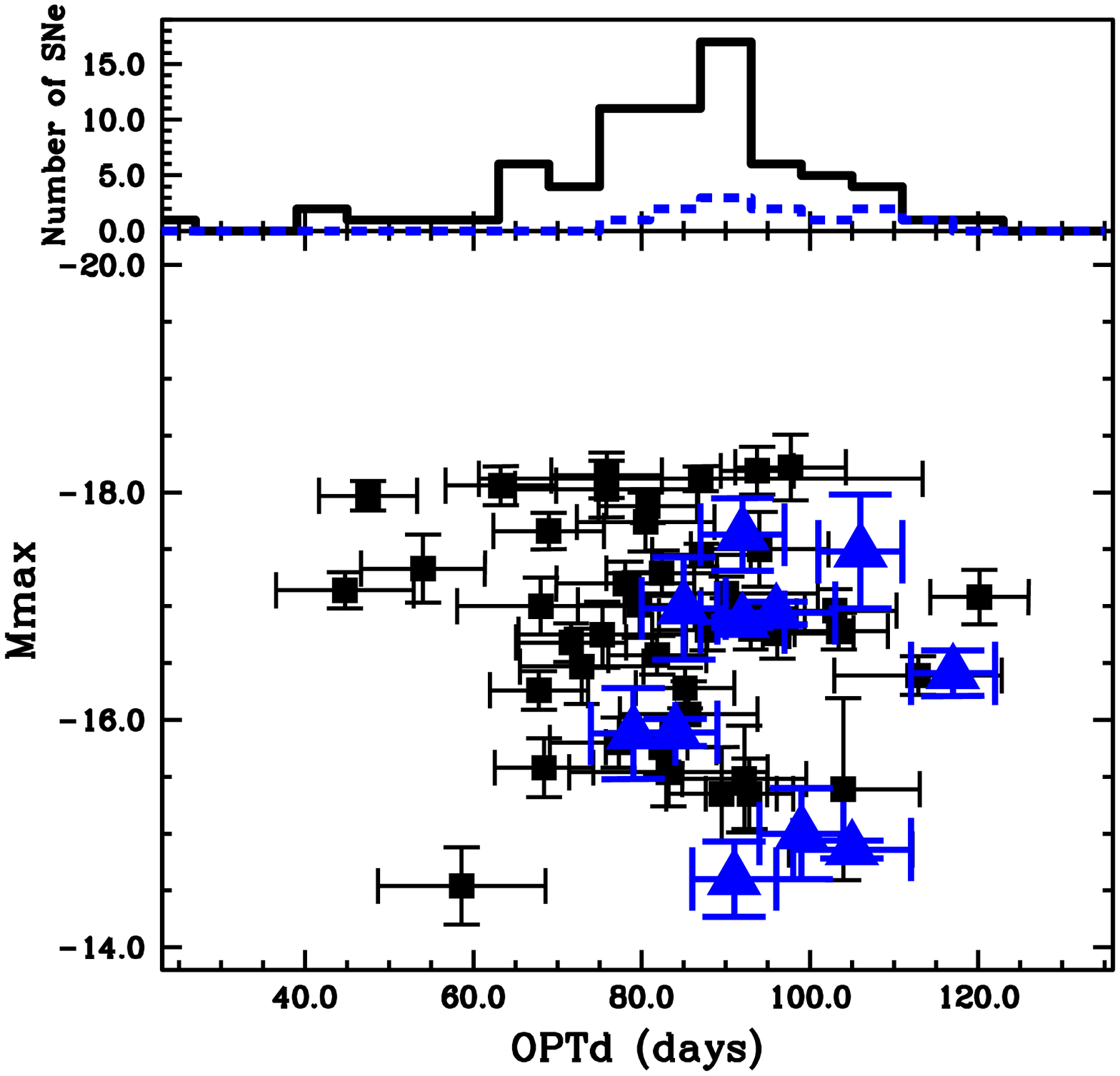} 
\caption{\textbf{\textit{Left, a)}}: SN~II absolute magnitudes at maximum ($M_{max}$)
plotted against `plateau' decline rate ($s_2$) \textbf{\textit{Right, b)}}: $M_{max}$
against optically thick phase durations (OPTds). Squares represent
the distribution from A14, while triangles show SNe~II with progenitor detection constraints. In 
the upper panel histograms for the two distributions of OPTd are also presented, solid: A14, 
and dashed: direct detection constraints.}
\end{figure}

\end{document}